# Tailoring Dirac fermions by *in-situ* tunable high-order moiré pattern in graphene-monolayer xenon heterostructure

Chunlong Wu[1§], Qiang Wan[1§], Cao Peng[1], Shangkun Mo[1], Renzhe Li[1], Keming Zhao[1], Yanping Guo[2], Shengjun Yuan[2], Fengcheng Wu[2,3*], Chendong Zhang[2,3] and Nan Xu[1,3*]

[1] *Institute of Advanced Studies, Wuhan University, Wuhan 430072, China*

[2] *School of Physics and Technology, Wuhan University, Wuhan 430072, China*

[3] *Wuhan Institute of Quantum Technology, Wuhan 430206, China*

§ There authors contributed equally.

* E-mail: nxu@whu.edu.cn; wufcheng@whu.edu.cn

A variety of novel quantum phases have been achieved in twist bilayer graphene (tBLG) and other moiré superlattices recently, including correlated insulators, superconductivity, magnetism, and topological states. These phenomena are very sensitive to the moiré superlattices, which can hardly be changed rapidly or intensely. Here, we report the experimental realization of a high-order moiré pattern (a high-order interference pattern) in graphene-monolayer xenon heterostructure (G/mXe), with moiré period *in-situ* tuned from few nanometers to $+\infty$ by changing the lattice constant of Xe through different annealing temperatures and pressures. We use angle-resolved photoemission spectroscopy to directly observe that replicas of graphene Dirac cone emerge and move close to each other in momentum-space as moiré pattern continuously expands in real-space. When the moiré period approaches $+\infty$, the replicas finally overlap with each other and an energy gap is observed at the Dirac point induced by intervalley coupling, which is a manifestation of Kekulé distortion. We construct a continuum moiré Hamiltonian, which can explain the experimental results well. The form of moiré Hamiltonian in G/mXe is similar to that in tBLG, and moiré band with narrow bandwidth is predicted in G/mXe. However, the moiré Hamiltonian couples Dirac fermions from different valleys in G/mXe, instead of ones from different layers in tBLG. Our work demonstrates a novel platform to study the continuous evolution of moiré pattern and its modulation effect on electronic structure, and provides an unprecedented approach for tailoring Dirac fermions with tunable intervalley coupling.

Moiré superlattices have become a promising playground for explorations of novel quantum states and phase transitions. In twisted bilayer graphene (tBLG), moiré pattern couples Dirac fermions from the same valley but different consisting layers (Fig. 1a), and results in moiré bands with extremely narrow bandwidth at the magic-angle condition [1]. Novel quantum phases, including correlated insulators [2-6], superconductivity [7-15], magnetism [16-19] and nontrivial topological states [20-22], are experimentally realized by designing moiré patterns and tuning the electron filling of moiré bands. The flat moiré bands in tBLG are directly observed by state-of-the-art angle-resolve photoemission spectroscopy (ARPES) with nanoscale real-space resolution [23-24]. A similar moiré modulation effect can also be realized in bilayers composed of other two-dimensional (2D) materials. For example, moiré pattern in twisted transition metal dichalcogenides (tTMD) also couple states from the same valley but different layers, which can induce flat moiré bands [25] and novel quantum states [26-28] like Mott insulators [29] and generalized Wigner crystal phase [30].

These novel quantum phenomena are very sensitive to the lattice structures of moiré patterns. Therefore, *in-situ* tunable moiré patterns can provide a powerful tool to design electronic structures, search novel quantum phases, and investigate phase transitions. The moiré pattern is determined by twist angle (θ) and lattice constant mismatch ($\Delta a = a_1 - a_2$) between the consisting layers. One way to realize tunable moiré pattern is to control the twist angle θ, which has been successfully achieved in elaborate devices [31-32]. Using an atomic force microscope (AFM) tip to push the specially shaped BN on graphene, θ can be changed in a single device and the moiré patterns with different periods can be realized. However, both fabrications of the special devices and control of θ using AFM tips are technically challenging. Moreover, although θ can be fully tuned from $0° - 360°$, the tunable range of moiré period is rather limited in a practical system (for instance, it's ~ 14 nm in graphene-BN [31]). Therefore, rapidly tuning moiré pattern in a large range of moiré period is still elusive.

Here, we report a new route to realize *in-situ* tunable high-order moiré pattern in graphene-monolayer xenon heterostructure, which we term as G/mXe, with moiré period tuned in a range from nanometer-scale to $+\infty$. Distinct from the previous method by changing twist angle θ, we manage to tune the lattice constant of Xe monolayer by annealing G/mXe at different temperatures and pressures above the second Xe layer desorption temperature, therewith to control the lattice mismatch term. Using lab-based ARPES, we directly observe the replicas of graphene Dirac cones generated by the moiré pattern, which move close to each other in momentum-space ($k$-space) as moiré pattern expands in real-space. The replicas finally overlap with each other as moiré period approaches $+\infty$ and moiré pattern evolves into a commensurate structure where a $\sqrt{3} \times \sqrt{3}$ superlattice with Kekulé distortion develops in graphene. A gap with $\Delta_G$ = 0.4 eV is then observed at Dirac point, induced by intervalley coupling of Dirac fermions. A continuum moiré Hamiltonian $\mathcal{H}(\boldsymbol{k},\boldsymbol{r})$ of G/mXe can explain experimental results qualitatively well. The $\mathcal{H}(\boldsymbol{k},\boldsymbol{r})$ in G/mXe has a form similar to that in tBLG; however, it couples Dirac fermions from different valleys in G/mXe (Fig. 1b) instead of from different graphene layers in tBLG (Fig. 1a). Nonetheless, moiré band with a narrow bandwidth is predicted in G/mXe model. Our work not only offers a promising platform for tailoring Dirac fermions with tunable intervalley coupling, but also provides a novel way to induce and tune the moiré pattern in 2D materials by noble gas monolayer.

Noble gas films deposited on metal surfaces have been intensively studied previously with a focus mainly on their lattice structures [33]. Xe monolayer has been grown on graphite surface [34], forming a hexagonal structure with primitive vectors $30°$ rotated from those of graphite. Our low-energy electron diffraction (Fig. S1 in Supplementary Materials) and ARPES results on G/mXe consistently indicate that Xe monolayer shows the same structure on graphene and graphite. In G/mXe (Fig. 1d), both $\Delta a = a_{Xe} - a_{Gra}$ and $\theta$ between the unit cells of graphene and Xe monolayer are large, where $\theta = 30°$, $a_{Gra} \approx 2.46$ Å and $a_{Xe} \sim 4.4$ Å are the lattice constants of graphene and Xe monolayer, respectively ($a_{Xe}$ is annealing process dependent as discussed in the following). Therefore, the first-order moiré effect directly related to unit cells of Xe monolayer and graphene is expected to be weak in G/mXe.

However, a $\sqrt{3} \times \sqrt{3}$ supercell of graphene and the unit cell of Xe are rotationally aligned ($\theta' = \theta - 30° = 0°$) and have a small lattice constant mismatch $\Delta a' = a_{Xe} - \sqrt{3}a_{Gra}$ (Fig. 1d and more discussion in Fig. S2 in Supplementary Materials). Therefore, the high-order moiré effect arising from $\Delta a'$ is expected to be pronounced with a moiré period given by

$$a'_m = \frac{a_{Gra} a_{Xe}}{\Delta a'} \quad (1).$$

The large-scale atomic arrangement plotted in Fig. 1c consistently indicates the high-order moiré pattern in G/mXe. Similar to tBLG, moiré pattern in G/mXe forms a hexagonal structure, with center and corner regions showing different stacking forms (Figs. 1d and e-f).

In momentum $k$-space, moiré pattern modulates electronic structure. Replicas of Dirac cones are induced by shifting the original Dirac cones of graphene with momentum transfers of the first-shell reciprocal lattice vectors of Xe monolayer $\vec{G}_{Xe}$. As illustrated in Fig. 1b, the moiré pattern induced replicas are located at the corners of a hexagon that is centered at the Γ point. This hexagon is the moiré Brillouin zone (BZ) corresponding to the high-order moiré pattern with a period $a'_m$ as discussed in Fig. S3 in Supplementary Materials.

In addition, replicas can be induced near the K (K') point also by scatterings of momentum $\vec{G}_{Xe}$. Two replicas and one original Dirac cone are found at three corners of a hexagon that is centered at the $\vec{G}_{Xe}$ point. This hexagon is a momentum-shifted copy of the moiré BZ centered at the Γ point. The replicas have the same valley arrangement as the original Dirac cones of graphene, i.e., the nearest neighbor Dirac cones have different chirality.

We perform ARPES measurements on G/mXe at T = 67 K and directly observe the replicas of graphene Dirac cone, as well as energy bands of both graphene and Xe monolayer. The graphene Dirac cone and its replicas appear near both the Γ and K points in Fermi surface (FS) mapping in Fig. 2a. In addition, the Xe monolayer bands are clearly observed at high binding energy ($E_B$) in Fig. 2b, in good agreement with band calculations of free-standing Xe monolayer [35]. Three dispersive bands are observed, corresponding to the Xe 5$p$ orbits. We resolve a spin-orbit coupling (SOC) induced energy gap ($\Delta_{SOC}$) at the Γ point between the $j = \frac{3}{2}$ bands with different $|m_j|$, with a value of $\Delta_{SOC}$ = 0.48 eV determined from the energy distribution curve (EDC) at the Γ point in Fig. 2b. There are additional nondispersive features, which are also observed previously for Xe monolayer on the metal surface [36]. They could be induced by partial Xe atoms without long-range ordering which are not crystallized well during the film depositing process.

We turn to describe the replicas of graphene Dirac cone near the Γ point, with the FS and constant energy plot at $E_B$ = 1.6 eV in a zoomed-in area shown in Figs. 2c and d, respectively. The band evolution is fully consistent with six replicas $\alpha_1$-$\alpha_6$ sitting at corners of moiré BZ, which are induced by moiré pattern as discussed previously. Figure 2e displays the band structure along cut 2 in Fig. 2c, with a pair of Dirac cone replicas $\alpha_1$ and $\alpha_4$ clearly observed; we determine the momentum center of the $\alpha_4$ replica to be $\boldsymbol{\kappa} = (0.09\ \text{Å}^{-1}, 0)$ at T = 67 K.

We also observe the replicas near the K point as indicated by Fig. 2a. In the constant energy plot at different $E_B$ in Fig. 2f-h, additional intensities are observed around the K point. We can clearly resolve the replicas $\beta_1$ and $\beta_2$ in band structure plots in Figs. 2i and 2j, which are along cut3 and cut4 in Fig. 2f, respectively. The band structure evolutions shown in Fig. 2f-j are fully consistent with a pair of Dirac cone replicas sitting at the corners of the moiré BZ near the K point.

Due to the moderate in-plane van der Waals (vdW) interaction between charge-neutral Xe atoms, we manage to continuously control $a_{Xe}$ with a variation of about 4% by annealing at different temperatures above the second-layer desorption temperature $T_s$. $T_s$ is Xe pressure ($P_{Xe}$) dependent with $T_s$ ~ 51K at $P_{Xe}$ = 1×10$^{-9}$ Torr. Therewith, we can tune moiré pattern in G/mXe by modifying the $\Delta a'$ term according to formula *(1)*. Such kind *in-situ* tuning of moiré pattern is experimentally demonstrated by the moving of replicas in *k*-space as a function of temperature. In Fig. 3a-b, we present temperature-dependent ARPES measurements above $T_s$ with $P_{Xe}$ = 1×10$^{-9}$ Torr. It is clearly seen that the distance between the $\alpha_1$ and $\alpha_4$ replicas (marked by red and blue arrows, respectively) can be directly tuned by temperature. The $\alpha_1$ and $\alpha_4$ replicas touch each other at $E_F$ (0.39 eV above the Dirac points) at T = 57.5 K (Fig. 3b-VI). As T decreasing, $\alpha_1$ and $\alpha_4$ replicas further move towards each other and they finally overlap and merge into a single Dirac cone replica at T ~ 51.5 K (Fig. 3b-VX). The FS also evolves from six circles surrounding the Γ point at T = 67 K (Fig. 2d) into a single circle at T = 51.5 K (Fig. 3c). We can experimentally determine the T-dependence of the replica's momentum $\boldsymbol{\kappa}$, and extract $a'_m$ and $a_{Xe}$, with the results shown in Figs. 3d-e, respectively. Note that the change in $a_{Gra}$ should be negligible in the temperature range of 50~70 K. The T-dependent measurements on the original Dirac cone at the K point consistently show no change of $a_{Gra}$, as demonstrated and discussed in Fig. 4. The extracted $|\kappa|$ shows a good T-linear relationship above 55 K (Fig. 3d). For T < 55 K, $\kappa$ is too small and beyond the *k*-resolution of ARPES experiments, but a linear extrapolation fits the experimental results in Fig. 3a-b well. Results in Fig. 2f demonstrate that ARPES measurement on moiré pattern induced replicas provides a sensitive method to determine lattice constant of Xe with the resolution of the order of 0.1 Å.

As $a_{Xe}$ approaches $\sqrt{3}a_{Gra}$ at low T (Fig. 3f), according to formula *(1)*, $\Delta a'$ approaches zero and $a'_m$ diverges in G/mXe (Fig. 3e). In this process, high-order moiré pattern in G/mXe with $a'_m \to +\infty$ evolves into a Kekulé distortion [37] with a $\sqrt{3} \times \sqrt{3}$ reconstruction of graphene. Replicas near the Γ point overlap with each other with $\kappa \to 0$ (Fig. 3b-VX and c). Because the nearest replicas have different valley chirality, Kekulé gap is expected as induced by intervalley coupling. The Kekulé gap is directly observed in EDC at the Γ point for $a'_m \to +\infty$ with $\Delta_G$ = 0.4 eV (Fig. 3b-VX). The gap value is similar as that observed in other Kekulé systems [38]. The observation of Kekulé gap excludes the photoelectrons surface diffraction as the reason of replicas in ARPES experiments.

The *in-situ* tunable moiré pattern in G/mXe offers an opportunity to study moiré modulation effect on Dirac fermion with variable $a'_m$. In Fig. 4a, we plot the band structure of G/mXe near the K point with different $a'_m$ moiré patterns. The original Dirac cone γ keeps at the K point for all the moiré patterns. Therefore, we plot EDCs at the K point for different $a'_m$ in Fig. 4b, as well as the one of pristine graphene for comparison. The EDC of $a'_m = 5$ nm moiré pattern shows a single peak feature, similar to EDC of pristine graphene. We also observe considerable spectra weight at $E_B$ > 1 eV, which corresponds to β$_2$ replicas as seen from Fig. 4a-I. For moiré pattern with larger $a'_m$ in Fig. 4a, $\kappa$ becomes smaller rapidly (as seen from Fig. 3d-e) and the replicas β$_1$ and β$_2$ move close to the γ cone. The spectra weight corresponding to replicas at high $E_B$ shifts towards the Dirac point, and leads to a broad peak with nearly flat top for $a'_m = 10$ and 15 nm in the EDC plot. As $a'_m$ further increasing, the EDC evolves into a double-peak shape. This double-peak feature remains as a Kekulé distortion form with $a'_m \to +\infty$, and an energy gap with $\Delta_G$ = 0.4 eV is observed at the K point. It is fully consistent with the Kekulé gap observed at the Γ point (Fig. 3b-VX).

To capture the experimental observation, we construct a continuum moiré Hamiltonian $\mathcal{H}$ for Dirac fermions in G/mXe based on symmetries, which is discussed in detail in the Supplementary Materials. The form of $\mathcal{H}$ is similar to that of tBLG, but it couples Dirac fermions from two valleys in graphene, instead of from two consisting layers in tBLG. The Hamiltonian $\mathcal{H}$ is fully consistent with the scenario discussed in Fig. 2a, and captures the main features of the ARPES results. For a finite $a'_m$, the Dirac cone replicas from two valleys are located at the moiré BZ corners $\kappa'$ and $\kappa$.

We calculate the theoretical ARPES spectra near the K point for different $a'_m$ and directly compare with the experimental results in Fig. 4a. As shown in the inset of Fig. 4a, hybridization gaps open at the crossing points between γ cone and replicas. Although the hybridization gaps cannot be resolved in ARPES results due to the energy broadening in the experimentally observed bands, there are some sudden intensity changes around the hybridization energies (arrows in Fig. 4a-II). Similar to tBLG systems, the Dirac nodes from the two valleys remain gapless for a finite $a'_m$ in the theoretical band structure because of $\hat{C}_{2z}\hat{T}$ symmetry, where $\hat{C}_{2z}$ is the twofold rotation symmetry around $z$ axis and $\hat{T}$ is time-reversal symmetry, and form a moiré band separated from other states. The bandwidth of this moiré band can be tuned by $a'_m$, and become very narrow for large $a'_m$ (Figs. 4a-V and Fig. S6 in Supplementary Materials). Given the moiré band with narrow bandwidth, correlated states and superconductivity could emerge in G/mXe.

Our model can also qualitatively explain the evolution of EDCs at the K point for different $a'_m$ in Fig. 4b. As plotted in Fig. 4c, two side peaks corresponding to the replicas show in the theoretical EDC for $a'_m$ = 10 nm. Although we cannot experimentally resolve the three-peak structure in Fig. 4b due to the energy broadening in ARPES experiments, the broader peak in EDC for $a'_m$ = 10 nm than that for $a'_m$ = 5 nm indicates two side-peaks close to each other. As $a'_m$ increases, the main peak at Dirac point becomes weaker and the two side peaks become stronger in theory of in Fig. 4c. With less and less weight of the main peak, we can experimentally resolve the two side-peaks for $a'_m > 20$ nm in Fig. 4b. In the limit of $a'_m \to +\infty$ (i.e., graphene and Xe become commensurate with $a_{Xe} = \sqrt{3} a_{Gra}$ and $\kappa \to 0$), the main peak at Dirac point is expected to vanish and the energy separation between the two side peaks defines the Kekulé distortion induced gap. However, the experimental peak width is on the same order as the double peak separation. Therefore, there are still finite spectra weight at the Dirac point in ARPES results for $a'_m \to +\infty$ in Fig. 4b.

The consistent experimental and theoretical results suggest that flat moiré bands can appear in G/mXe for a large moiré period $a'_m$, which can be realized in a wide range of temperature as discussed in the following. When the temperature is above the second-layer desorption temperature $T_s$, $a'_m$ in G/mXe can be directly tuned by Xe pressure $P_{Xe}$ (Fig. S4 in Supplementary Information). Moiré pattern with the same $a'_m$ can be realized at higher temperature by increasing Xe pressure. Therefore, moiré pattern in G/mXe can be tuned under moderate conditions without the requirement of high-vacuum.

When the temperature is below $T_s$, we can also tune the high-order moiré pattern in G/mXe through different annealing processes. As shown in Fig. S5 Supplementary Information, we can tune $a'_m$ within a large range at T = 6 K, that is the lowest T for our ARPES facility. Therefore, low temperature transport and spectroscopy measurements on G/mXe are feasible to resolve the possible flat moiré band and related quantum phases.

In summary, we demonstrate an *in-situ* tunable high-order moiré pattern with $a'_m$ from nanometer scale to $+\infty$ in G/mXe system. It is realized in a novel way by controlling the $\Delta a'$ term between the primitive vectors of graphene $\sqrt{3} \times \sqrt{3}$ supercell and Xe monolayer unit cell, by different annealing processes. The moiré pattern induced Dirac cone replicas are directly observed by ARPES measurements, which can move in *k*-space as $a'_m$ changes. Moiré pattern evolves into a Kekulé distortion as $a'_m \to +\infty$, and replicas overlap with each other with a gap opened at Dirac point. The theoretical model indicates flat band physics hosted in G/mXe. Our results demonstrate that G/mXe provides an opportunity for ARPES probes of moiré physics without the need of nanoscale spatial resolution. However, the band structures of both bare graphene and G/mXe observed by lab-based ARPES facility here are broader than that of graphene and tBLG observed by synchrotron-based ARPES [23-24], and we cannot resolve fine features of moiré bands . The possible reasons are the matrix element effect of incident light from He-lamp. The synchrotron-based ARPES with tunable photon energy could be helpful to resolve fine features near the Dirac point for different $a'_m$. Transport measurements with gate tunable electron filling would provide evidence for possible correlated states, superconductivity and other flat band related quantum states in G/mXe. The high-order moiré pattern for systems with large lattice mismatch $\Delta a$ and twist angle $\theta$ was also observed in graphene-SiC system [39]. We have realized graphene- and bilayer-graphene-based heterostructures consisting of other noble gas monolayers e.g. krypton (Kr) and argon (Ar), and the *in-situ* tunable high-order moiré pattern is generally observed in these systems. We also observed similar behaviors in heterostructure consisting of layered vdW superconductor $Cu_xTiSe_2$ and monolayer Kr/Xe. Our work opens up vast new opportunities provided by the *in-situ* tunable moiré patterns in designing quantum phases of matter in 2D materials.

## References

[1] R. Bistritzer, and A. H. MacDonald, *Moiré bands in twist double-layer graphene.* PNAS **26**,108 (2011).

[2] Y. Cao, V. Fatemi, A. Demir, S. A. Fang, S. L. Tomarken, J. Y. Luo, J. D. Sanchez-Yamagishi, K. J. Watanabe, T. Taniguchi, E. Kaxiras, R. C. Ashoori, and P. Jarillo-Herrero, *Correlated insulator behaviour at half- lling in magic-angle graphene superlattices.* Nature **556**, 80 (2018).

[3] G. R. Chen, L. L. Jiang, S. Wu, B. Lyu, H. Y. Li, B. L. Chittari, K.Watanabe, T. Taniguchi, Z. W. Shi, J. Jung, Y. B. Zhang, and F. Wang, *Evidence of a gate-tunable Mott insulator in trilayer graphene moiré superlattice.* Nat. Phys. **15**, 237 (2019).

[4] G. R. Chen, A. L. Sharpe, P. Gallagher, I. T. Rosen, E. J. Fox, L. L. Jiang, B. Lyu, H. Y. Li, K. Watanabe, T taniguchi, J. Jung, Z. W. Shi, D. Goldhaber-Gordon, Y. B. Zhang, and F. Wang, *Signatures of tunable superconductivity in a trilayer graphene moiré superlattice.* Nature **572**, 215 (2019).

[5] G. W. Burg, J. H. Zhu, T. Taniguchi, K. Watanabe, A. H. MacDonald, and E. Tutuc,

*Correlated insulating states in twisted double bilayer graphene*. Phys. Rev. Lett. **123**, 197702 (2019).

[6] C. Shen, Y. B. Chu, Q. S. Wu, N. Li, S. P. Wang, Y. C. Zhao, J. Tang, J. Y. Liu, J. P. Tian, K. Watanabe, T. Taniguchi, R. Yang, Z. Y. Meng, D. X. Shi, O. V. Yazyev, and G. Y. Zhang, *Correlated states in twisted double bilayer graphene*. Nat. Phys. **16**, 520 (2020).

[7] Y. Cao, V. Fatemi, S. A. Fang, K. J. Watanabe, T. Taniguchi, E. Kaxiras, and P. Jarillo-Herrero, *Unconventional superconductivity in magic-angle graphene superlattices*. Nature **556**, 43 (2018).

[8] X. B. Lu, P. Stepanov, W. Yang, M. Xie, M. A. Aamir, I. Das, C. Urgell, K. Watanabe, T. Taniguchi, G. Y. Zhang, A. Bachtold, A. H. MacDonald, and D. K. Efetov, *Superconductors, orbital magnets and correlated states in magic angle bilayer graphene*. Nature **574**, 653 (2019).

[9] E. Codecido, Q. Y. Wang, R. Koester, S. Che, H. D. Tian, R. Lv, S. Tran, K. Watanabe, T, Taniguchi, F. Zhang, M. Bockrath, and C. N. Lau. *Correlated insulating and superconducting states in twisted bilayer graphene below the magic angle*. Sci. Adv. **5**, eaaw9770 (2019).

[10] Y. Saito, J. Y. Ge, K. Watanabe, T. Taniguchi, and A. F. Young, *Independent superconductors and correlated insulators in twisted bilayer graphene*. Nat. Phys. **16**, 926 (2020).

[11] P. Stepanov, I. Das, X. B. Lu, A. Fahimniya, K. Watanabe, T. Taniguchi, F. H. L. Koppens, J. Lischner, L. Levitov, and D. K. Efetov, *Untying the insulating and superconducting orders in magic-angle graphene*. Nature **583**, 375 (2020).

[12] M. Yankowitz, S. W. Chen, H. Polshyn, Y. X. Zhang, K. Watanabe, T. Taniguchi, D. Graf, A. F. Young, and C. R. Dean, *Tuning superconductivity in twisted bilayer graphene*. Science **363**, 1059 (2019).

[13] H. S. Arora, R. Polski, Y. R. Zhang, A. Thomson, Y. Choi, H. Kim, Z. Lin, I. Z. Wilson, X. D Xu, J. -H. Chu, K. Watanabe, T. Taniguchi, J. Alicea, and S. Nadj-Perge, *Superconductivity in metallic twisted bilayer graphene stabilized by $WSe_2$*. Nature **583**, 379 (2020).

[14] G. R. Chen, L. L. Jiang, S. Wu, B. Lyu, H. Y. Li, B. L. Chittari, K. Watanabe, T. Taniguchi, Z. W. Shi, J. Jung, Y. B. Zhang, and F. Wang, *Evidence of a gate-tunable Mott insulator in trilayer graphene moiré superlattice*. Nat. Phys. **15**, 237 (2019).

[15] G. R. Chen, A. L. Sharpe, P. Gallagher, I. T. Rosen, E. J. Fox, L. L. Jiang, B. Lyu, H. Y. Li, K. Watanabe, T taniguchi, J. Jung, Z. W. Shi, D. Goldhaber-Gordon, Y. B. Zhang, and F. Wang, *Signatures of tunable superconductivity in a trilayer graphene moiré superlattice*. Nature **572**, 215 (2019).

[16] A. L. Sharpe, E. J. Fox, A. W. Barnard, J. Finney, K. Watanabe, T. Taniguchi, M. A. Kastner, and D. Goldhaber-Gordon, *Emergent ferromagnetism near three-quarters lling in twisted bilayer graphene*. Science **365**, 605 (2019).

[17] Y. Cao, D. Rodan-Legrain, O. Rubies-Bigorda, J. M. Park, K. Watanabe, T. Taniguchi, and P. Jarillo-Herrero, *Tunable correlated states and spin-polarized phases in twisted bilayer-bilayer graphene*. Nature **583**, 215 (2020).

[18] X. M. Liu, Z. Y. Hao, E. Khalaf, J. Y. Lee, Y. Ronen, H. Yoo, D. H. Najafabadi, K. Watanabe, T. Taniguchi, A. Vishwanath, and P. Kim, *Tunable spin-polarized correlated states in twisted double bilayer graphene*. Nature **583**, 221(2020).

[19] M. H. He, Y. H. Li, J. Q. Cai, Y. Liu, K. Watanabe, T. Taniguchi, X. D. Xu, and M. Yankowitz, *Symmetry breaking in twisted double bilayer graphene*. Nat. Phys. **17**, 26 (2020).

[20] M. Serlin, C. L. Tschirhart, H. Polshyn, Y. Zhang, J. Zhu, K. Watanabe, T. Taniguchi, L. Balents, and A. F. Young, *Intrinsic quantized anomalous Hall e ect in a moiré heterostructure*. Science **367**, 900 (2020).

[21] S. W. Chen, M. H. He, Y. H. Zhang, V. Hsieh, Z. Y. Fei, K. Watanabe, T. Taniguchi, D. H. Cobden, X. D. Xu, C. R. Dean, and M. Yankowitz, *Electrically tunable correlated and topological states in twisted monolayer–bilayer graphene*. Nat. Phys. **17**, 374 (2021).

[22] G. Chen, A. L. Sharpe, E. J. Fox, Y. -H. Zhang, S. X. Wang, L. L. Jiang, B. Lyu, H. Y. Li, K. J. Watanabe, T. Taniguchi, Z. W. Shi, T. Senthil, D. Goldhaber-Gordon, Y. B Zhang, and F. Wang, *Tunable correlated Chern insulator and ferromagnetism in a moiré superlattice*. Nature **579**, 56 (2020).

[23] M. I.B. Utama, R. J. Koch, K. Lee, N. Leconte, H. Y. Li, S. Zhao, L. L. Jiang, J. Y. Zhu, K. Watanabe, T. Taniguchi, P. D. Ashby, A. Weber-Bargioni, A. Zettl, C. Jozwiak, J. Jung, E. Rotenberg, A. Bostwick, and F. Wang, *Visualization of the flat electronic band in twisted bilayer graphene near the magic angle twist.* Nat. Phys. **17,** 184 (2021).

[24] S. Lisi, X. B. Lu, T. Benschop, T. A. D. Jong, P. Stepanov, J. R. Duran, F. Margot,

I. Cucchi, E. Cappelli, A. Hunter, A. Tamai, V. Kandyba, A. Giampietri, A. Barinov, J. Jobst, V. Stalman, M. Leeuwenhoek, K. Watanabe, T. Taniguchi, L. Rademaker, S. J. V. D. Molen, M. P. Allan, D. K, Efetov, and F. Baumberger, *Observation of flat bands in twisted bilayer graphene*. Nat. Phys. **17,** 189 (2021).

[25] F. C. Wu, T. Lovorn, E. Tutuc, and A. H. MacDonald, *Hubbard model physics in transition metal dichalcogenide moiré bands*. Phys. Rev. Lett. **121**, 026402 (2018).

[26] E. C. Regan, D. Q. Wang, C. H. Jin, M. B. Utama, B. N. Gao, X. Wei, S. H. Zhao, W. Y. Zhao, Z. C. Zhang, K. Yumigeta, M. Blei, J. D. Calström, K. J. Watanabe, T. Taniguchi, S. Tongay, M. Crommie, A. Zettl, and F. Wang, *Mott and generalized Wigner crystal states in WSe2/WS2 moiré superlattices*. Nature **579**, 359 (2020).

[27] Y. Tang, L. Z. Li, T. X. Li, Y. Xu, S. Liu, K. Barmak, K. Watanabe, T. Taniguchi, A. H. MacDonald, J. Shan, and K. F. Mak, *Simulation of Hubbard model physics in $WSe_2/WS_2$ moiré superlattices*. Nature **579**, 353 (2020).

[28] Y. Xu, S. Liu, D. A. Rhodes, K. Watanabe, T. Taniguchi, J. Hone, V. Elser, K. F. Mak, and J. Shan, *Correlated insulating states at fractional fillings of moiré superlattices*. Nature **587**, 214 (2020).

[29] L. Wang, E. Shin, A. Ghiotto, L. Xian, D. A. Rhodes, C. Tan, M. Claassen, D. M. Kennes, Y. S. Bai, B. Kim, K. Wantanabe, T. Taniguchi, X. Y. Zhou, J. Hone, A. Rubio, A. N. Pasupathy, and C. R. Dean, *Correlated electronic phases in twisted bilayer transition metal dichalcogenides*. Nat. Mater. **19**, 861(2020).

[30] H. Y. Li, S. W. Li, E. C. Regan, D. Q. Wang, W. Y. Zhao, S. Kahn, K. Yumigeta, M. Blei, T. Taniguchi, K. Watanabe, S. Tongay, A. Zettl, M. F. Crommie, and F. Wang, *Imaging two-dimensional generalized Wigner crystals*. Nature **597**, 650 (2021).

[31] R. Ribeiro-Palau, C. J. Zhang, K. Watanabe, T. Taniguchi, J. Hone, amd C. R. Dean, *Twistable electronics with dynamically rotatable heterostructures*. Science **361**, 690 (2018).

[32] N. R. Finney, M. Yankowitz, L. Muraleetharan, K. Watanabe, T. Taniguchi, C. R. Dean, and J. Hone. *Tunable crystal symmetry in graphene–boron nitride heterostructures with coexisting moiré superlattices*. Nat. Nano. **14**, 1029 (2019).

[33] S. Ishi, B. Viswanathan, *Adsorption of xenon atoms on metal surfaces*. Thin Solid Films **201**, 373 (1991).

[34] M. Hamichi, A. Q. D. Faisal, J. A. Venables, and R. Kariotis, *Lattice parameter and orientation of xenon on graphite at low pressures*. Phys. Rev. B **39**, 415 (1989).

[35] K. Hermann, J. Noffke, and K. Horn *Lateral interactions in rare gas monolayers: Band-structure models and photoemission experiments.* Phys. Rev. B **22**, 1022 (1980).

[36] T. Mandel, G. Kaindl, M. Demke, W. Fischer, and W. D. Schneider, *Layer-by-Layer Band Structure of Physisorbed Xe on Al(111).* Phys. Rev. Lett. **55,** 1638 (1984).

[37] C. Y. Chang, C. Chamon, and C. Mudry, *Electron fractionalization in two-dimensional graphenelike structures*, Phys. Rev. Lett. **98**, 186809 (2007).

[38] C. H. Bao, H. Y. Zhang, T. Zhang, X. Wu, L. P. Luo, S. H. Zhou, Q. Li, Y. H. Hou, W. Yao, L. W. Liu, P. Yu, J. Li, W. H. Duan, H. Yao, Y. L. Wang, and S. Y. Zhou, *Experimental Evidence of Chiral Symmetry Breaking in Kekulé-Ordered Graphene.* Phys. Rev. Lett. **126**, 206804 (2021).

[39] C. L. Wu, Q. Wan, C. Peng, S. K. Mo, R. Z. Li, K. M. Zhao, Y. P. Guo, C. D. Zhang, and N. Xu, *Observation of high-order moiré effect and multiple Dirac fermions replicas in graphene-SiC heterostructure.* Phys. Rev. B **104**, 235130 (2021).

**Method**

Monolayer graphene was achieved by high-temperature annealing process [40] of n-type 6H SiC(0001) from PrMat. The monolayer graphene samples were characterized and conformed by scanning tunneling microscopy measurements. Xe monolayers were formed on clean graphene surface by depositing slightly thicker layers at T = 44 K with Xe pressure of $1\times10^{-9}$ Torr and then annealing at a temperature above the second-layer desorption temperature. ARPES measurements were performed at the home-designed facility with photon energy hν = 21.2 eV and photon spot size about 0.5 mm.

[40] Q. Wang *et al.*, *Large-scale uniform bilayer graphene prepared by vacuum graphitization of 6H-SiC (0001) substrates.* J. Phys.: Condens. Matter **25**, 095002 (2013).

# Figures

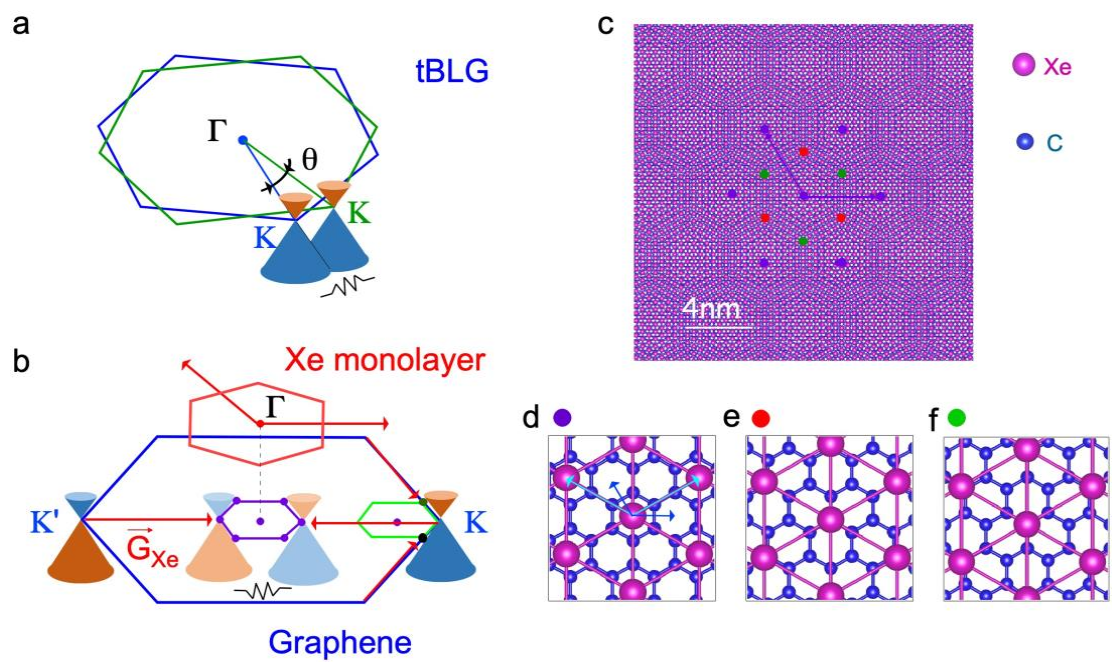

**Figure 1. High-order moiré pattern in G/mXe. a,** Schematic first-order moiré effect in tBLG systems in *k*-space. **b,** Same as **a**, but for high-order moiré effect in G/mXe. **c,** Large-scale atom arrangements showing moiré pattern of G/mXe in real-space. **d-f,** Atom arrangements in zoomed-in areas at the center and neighboring corners of moiré pattern, respectively. The primitive vectors of graphene and Xe monolayer are labeled in **d**.

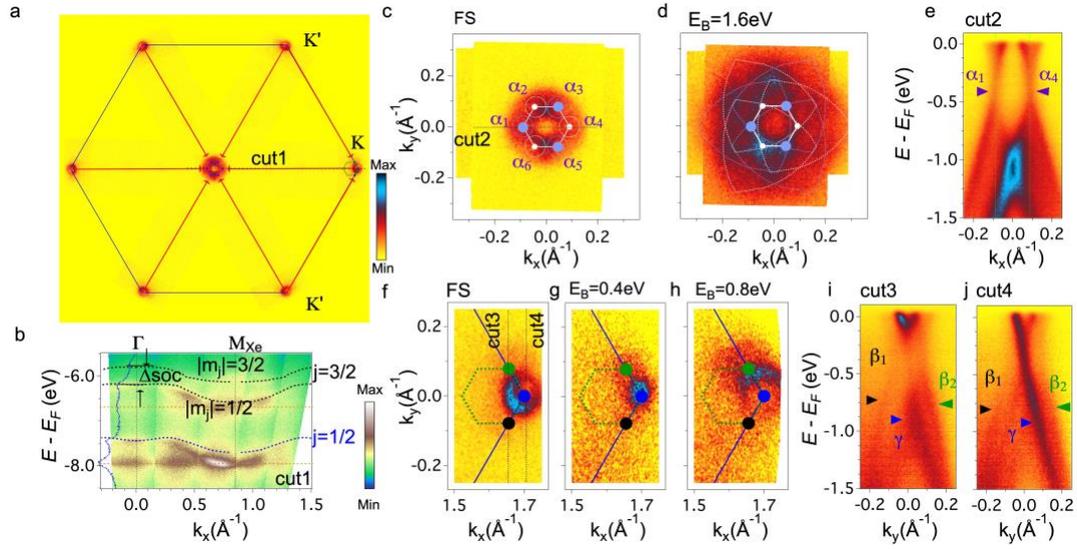

**Figure 2. Moiré modulation on band structure in G/mXe for T = 67 K. a,** Fermi surface mapping of G/mXe. **b,** Photoemission intensity plot at high $E_B$ along Γ-K direction, as indicated by cut1 in **a**. The calculation of free-standing Xe monolayer (red dashed lines) and EDC at Γ point are appended. **c-d,** Constant energy intensity plots near the Γ point at $E_F$ and $E_B$ = 1.6 eV, respectively. **e,** The band structure along the cut2 in **d**. **f-h,** Evolution of constant energy intensity plots near the K point. **i-j,** Photoemission intensity plots along cut3 and cut4 labeled in **f**, respectively.

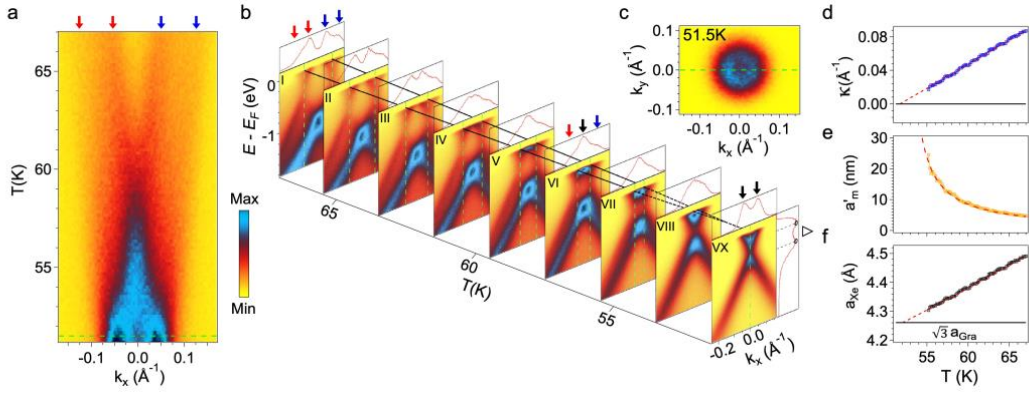

**Figure 3. The tunable moiré pattern in G/mXe. a,** T-dependent photoemission data at $E_F$ along the Γ-K direction. **b,** Corresponding T-dependent band structure. The MDCs at $E_F$ are also plotted on the top. The red and blue arrows indicate the Fermi momenta of the $\alpha_1$ and $\alpha_4$ replicas, respectively. **c,** The FS mapping near the Γ point at T = 51.5 K. **d-f,** T-dependent data of $\kappa$, $a'_m$ and $a_{Xe}$, respectively. The dashed lines are fitting curves. The value of $\sqrt{3}a_{Gra}$ is also indicated by the solid line in **f**.

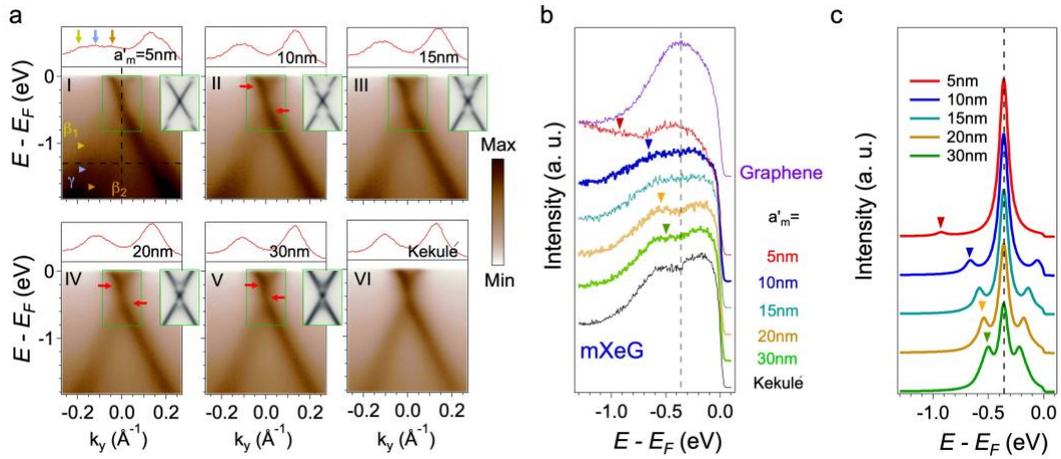

**Figure 4. Tailoring Dirac fermions in G/mXe. a,** The band structure evolution near the K point, with different moiré period $a'_m$. The MDC at $E_B = 1.3$ eV are appended on top, with colored arrows guiding the $\beta_1$ and $\beta_2$ replicas and original $\gamma$ cone. The simulated band structures for finite $a'_m$ are appended with the same energy- and k-scales for a direct comparison. **b,** The $a'_m$-dependent EDCs at the K point of G/mXe. The EDC of bare graphene is also plotted as purple line on the top. **c,** The calculated EDCs at the K point of G/mXe for finite $a'_m$. Different curves are shifted vertically for clarity.